УДК 551.466

# РАСПРОСТРАНЕНИЕ ВОЛНЫ КОНЕЧНОЙ АМПЛИТУДЫ В СТРАТИФИЦИРОВАННОЙ ЖИДКОСТИ ПЕРЕМЕННОЙ ГЛУБИНЫ

Диденкулова И.И.[1,2], Талипова Т.Г.[3], Пелиновский Е.Н.[1,3], Куркина О.Е.[1,4], , Родин А.А.[1,2], Панкратов А.С.[1], Наумов А.А.[1], Гиниятуллин А.Р.[1], Николкина И.Ф.[2]

[1] *Нижегородский государственный технический университет, Н. Новгород, Россия*
[2] *Институт Кибернетики, Таллинский технический университет, Таллин, Эстония*
[3] *Институт прикладной физики РАН, Н. Новгород, Россия*
[4] *Высшая школа экономики, Н. Новгород, Россия*

**АННОТАЦИЯ**

Приведена физико-математическая модель трансформации двумерных нелинейных внутренних волн в природных стратифицированных водоемах. Модель основана на уравнении Кортевега-де Вриза и его обобщениях. Коэффициенты этого уравнения вычисляются по известным вертикальным распределениям плотности жидкости с помощью решения задачи Штурма-Лиувилля на собственные значения. Модель учитывает горизонтальную изменчивость гидрологических полей по горизонтали и переменность глубины водоема. В качестве примера рассмотрена трансформация уединенной волны (солитона) в двухслойном потоке с уменьшающейся глубиной. Показано, что солитон разрушается в двух критических точках. Одна из них связана с переходом двухслойного потока в однослойный, а вторая – когда толщины слоев сравниваются между собой. Рассчитано изменение амплитуды уединенной волны от толщины нижнего слоя.

**ВВЕДЕНИЕ**

Нелинейные внутренние волны являются неотъемлемой частью волновых движений в стратифицированных средах, и они часто наблюдаются в океанах и морях. Эти волны могут достигать очень больших амплитуд ввиду малой разницы в плотностях соседних слоев, так что архимедовы силы, действующие на частицу жидкости, достаточно малы, чтобы воспрепятствовать передвижению частиц по вертикали. Общие сведения о внутренних волнах в океанах и морях можно найти в книгах (Морозов, 1985; Коняев и Сабинин, 1992; Hutter, 2012). Теоретические модели описания нелинейных внутренних волн в стратифицированной жидкости основаны на уравнениях Эйлера (если вязкостью жидкости пренебрегается) или уравнениях Навье-Стокса с различными аппроксимациями на диссипацию волновой энергии и эти модели описаны, например, в книгах (Миропольский, 1981; Vlasenko et al., 2005). Особо необходимо отметить, что в реальных природных водоемах вертикальная стратификация плотности воды не является фиксированной и меняется в пространстве. В настоящее время имеются достаточно подробные гидрологические атласы, позволяющие количественно оценивать характеристики вертикальной и горизонтальной изменчивости поля плотности (Boyer et al., 2006).

В течение последних двадцати лет в Институте прикладной физики РАН и в Нижегородском государственном техническом университете развивается физико-математическая модель описания трансформации нелинейных внутренних волн в стратифицированной жидкости с учетом распределения поля плотности в пространстве и переменности глубины жидкости (Пелиновский и др., 1977, 1994; Holloway et al., 1997, 1999; Талипова и др., 1999; Pelinovsky et al., 2007). Она основана на обобщенном уравнении Кортевега -де Вриза, являющимся в настоящее время эталонным уравнением нелинейной математической физики для описания нелинейных волновых процессов в слабодиспергирующих средах. Наиболее важными особенностями развитой модели является ее применимость к реальным гидрологическим полям без использования приближения многослойности вод океана или непрерывности вертикального профиля плотности воды. Эта модель сейчас реализована в виде вычислительного комплекса, позволяющего автоматизировать вычисления (Тюгин и др., 2011).

В настоящей работе с помощью данной модели изучается трансформация внутренней волны конечной амплитуды в двухслойном потоке переменной глубины.

## ТЕОРЕТИЧЕСКАЯ МОДЕЛЬ, ОСНОВАННАЯ НА ОБОБЩЕННОМ УРАВНЕНИИ КОРТЕВЕГА – ДЕ ВРИЗА

Наиболее строго, обобщенное (иногда говорят расширенное) уравнение Кортевега – де Вриза с использованием асимптотической процедуры выведено в работе (Пелиновский и др., 2000; Grimshaw et al., 2002). Здесь мы приведем основные уравнения модели в приближении Буссинеска (когда изменения плотности по вертикали достаточно малы – типичное приближение для природных водоемов) в пренебрежении вязкими процессами, а также не будем учитывать фоновые течения. В частности, если нет изменения характеристик среды по горизонтали, основным уравнением для двумерного волнового потока в стратифицированной жидкости является 1+1 уравнение Гарднера или уравнение Кортевега-де Вриза со смешанной (квадратично-кубической) нелинейностью

$$\frac{\partial \eta}{\partial t} + (\alpha\eta + \alpha_1\eta^2)\frac{\partial \eta}{\partial x} + \beta\frac{\partial^3 \eta}{\partial x^3} = 0, \quad (1)$$

где $\eta(x,t)$ – волновая функция, входящая в вертикальное смещение изопикн (поверхности равной плотности $\zeta(x,y,t)$).

$$\zeta(x,y,t) = \eta(x,t)\Phi(y). \quad (2)$$

Здесь $x$ – горизонтальная и $y$ – вертикальная координаты, $t$ – время, $\Phi(y)$ – модовая функция, являющаяся решением задачи Штурма-Лиувилля с нулевыми граничными условиями на собственные значения

$$\frac{d^2\Phi}{dy^2} + \frac{N^2(y)}{c^2}\Phi = 0, \quad \Phi(0) = \Phi(-H) = 0, \quad (3)$$

где $H$ – глубина слоя воды (отсюда видно, что начало координат связывается с водной поверхностью), $N(y)$ – частота Вяйсяля-Брента (плавучести), определяемая через вертикальное распределение плотности жидкости $\rho_0(y)$

$$N(y) = \sqrt{-\frac{g d\rho_0}{\rho_0 dy}}, \quad (4)$$

$g$ – ускорение силы тяжести и $c$ – собственное значение задачи Штурма – Лиувилля. Легко показать, что краевая задача (3) имеет дискретный спектр различающихся собственных значений (Миропольский, 1981), на чем мы не останавливаемся. В рамках данной модели каждая мода внутренних волн распространяется независимо, так что мы не будем использовать индекс для выбора определенной моды. Напомним, что собственное значение $c$ определяет физически скорость распространения длинных внутренних волн малой амплитуды. Фактически, уравнение (1) записано в системе отсчета, движущейся со скоростью c, так что координата x есть $x - ct$. Отметим также, что мы используем условие нормировки на функцию $\Phi(y)$: $\Phi_{max} = 1$, так что функция $\eta(x,t)$ описывает вертикальное смещение изопикны в максимуме моды.

Коэффициенты уравнения (1) называются соответственно коэффициентами квадратичной и кубической нелинейности и дисперсии; они определяются через вертикальное распределение плотности воды

$$\alpha = \left(\frac{3c}{2}\right) \frac{\int_{-H}^{0}(d\Phi/dy)^3 dy}{\int_{-H}^{0}((d\Phi/dy)^2 dy} \quad (5)$$

$$\alpha_1 = \frac{3c}{2}\frac{\int[3(dT/dy) - 2(d\Phi/dy)^2](d\Phi/dy)^2 dy}{\int(d\Phi/dy)^2 dy} - \frac{3c}{2}\frac{\int\{\alpha^2(d\Phi/dy)^2 - \alpha[5(d\Phi/dy)^2 - 4dT/dy]d\Phi/dy\}dy}{\int(d\Phi/dy)^2 dy}, \quad (6)$$

$$\beta = \left(\frac{c}{2}\right)\frac{\int_{-H}^{0}\Phi^2 dy}{\int_{-H}^{0}(d\Phi/dy)^2 dy}. \quad (7)$$

Здесь функция $T$ есть нелинейная поправка к модальной функции $\Phi$, определяемая уравнением

$$\frac{d^2T}{dy^2} + \frac{N^2(y)}{c^2}T = -\frac{\alpha}{c}\frac{d^2\Phi}{dy^2} + \frac{3}{2}\frac{d}{dy}\left[\left(\frac{d\Phi}{dy}\right)^2\right]. (8)$$

Уравнения модели достаточно громоздки, однако вычисление всех коэффициентов по реально заданному распределению плотности воды в настоящее время не представляет трудностей, и оно автоматизировано, например, в вычислительном комплексе (Тюгин и др., 2011).

На практике, как уже говорилось, необходимо учитывать переменность глубины бассейна, а также изменчивость плотности по горизонтали. Если изменения по трассе достаточно плавные, так что можно пренебречь отражением волновой энергии, то аналогично (1) можно вывести уравнение Гарднера с переменными коэффициентами (Holloway et al., 1997, 1999)

$$\frac{\partial \xi}{\partial x} + \left(\frac{\alpha(x)Q(x)}{c^2(x)}\xi + \frac{\alpha_1(x)Q^2(x)}{c^2(x)}\xi^2\right)\frac{\partial \xi}{\partial s} + \frac{\beta(x)}{c^4(x)}\frac{\partial^3 \xi}{\partial s^3} = 0, \quad (9)$$

где

$$\eta(s,x) = Q(x)\xi(s,x), \quad (10)$$

$$Q(x) = \sqrt{\frac{(Mc^3)_0}{Mc^3}}, \quad M(x) = \int_{-H(x)}^{0}\left(\frac{d\Phi}{dy}\right)^2 dy. \quad (11)$$

$$s(x,t) = \int \frac{dx}{c(x)} - t. \quad (12)$$

Здесь индекс 0 в $Q(x)$ соответствует начальному значению параметров.

Таким образом, разработанная модель позволяет учесть реальное распределение плотности воды по вертикали и горизонтали, а также переменность глубины бассейна. При этом задача Штурма-Лиувилля (3) должна решаться в каждой точке по трассе распространения, в результате все коэффициенты уравнения Гарднера становятся переменными по горизонтали. Именно поэтому уравнение (9) и называется уравнением Гарднера с переменными коэффициентами. Наконец важно отметить, что уравнение (9) имеет два закона сохранения потоков массы и энергии

$$\int_{-\infty}^{+\infty} \xi(s,x)ds = \int_{-\infty}^{+\infty} \xi(t,x)dt, \quad (13)$$

$$\int_{-\infty}^{+\infty} \xi^2(s,x)ds = \int_{-\infty}^{+\infty} \xi^2(t,x)dt, \quad (14)$$

которые могут быть использованы для контроля вычислений.

В данной работе будут исследованы волны в двухслойном потоке. В этом случае все коэффициенты вычисляются явно (Kakutani and Yamasaki, 1978)

$$c = \sqrt{\frac{g\Delta\rho}{\rho}\frac{h_1 h_2}{h_1 + h_2}}, \quad (15)$$

$$\alpha = \frac{3c}{2}\frac{h_1 - h_2}{h_1 h_2}, \quad (16)$$

$$\alpha_1 = -\frac{3c}{8h_1^2 h_2^2}(h_1^2 + h_2^2 + 6h_1 h_2), \quad (17)$$

$$\beta = \frac{ch_1 h_2}{6}, \quad (18)$$

где $h_1$ – толщина верхнего слоя, $h_2(x)$ – переменная толщина нижнего слоя, $\Delta\rho/\rho$ - скачок плотности на границе раздела слоев. Как видим, все коэффициенты знакоопределенные, за исключением коэффициента квадратичной нелинейности, который положителен, если граница раздела слоев находится ближе ко дну, и отрицателен в противоположном случае. Аналогично вычисляется коэффициент усиления

$$Q(x) = \sqrt{\frac{c_0^3}{c^3(x)}\frac{h_1^{-1} + h_2^{-1}(0)}{h_1^{-1} + h_2^{-1}(x)}}. \quad (19)$$

## УЕДИНЕННЫЕ ВНУТРЕННИЕ ВОЛНЫ (СОЛИТОНЫ)

Уравнение Гарднера (1) имеет стационарные решения в виде уединенных волн (солитонов)

$$\eta(x,t) = \frac{A}{1 + B\text{ch}(\gamma(x - Vt))}, \quad (20)$$

$$A = \frac{6\beta\gamma^2}{\alpha}, \quad B^2 = 1 + \frac{6\alpha_1\beta\gamma^2}{\alpha^2}, \quad V = \beta\gamma^2, \quad (21)$$

где $\gamma$ есть произвольный параметр, характеризующий обратную ширину солитона. Другим свободным параметром является фаза солитона (его местоположение), которую мы здесь не рассматриваем. Амплитуда волны, или экстремум функции (20), равна

$$a = \frac{A}{1 + B}. \quad (22)$$

*Солитон Кортевега – де Вриза.* В случае, когда кубическая нелинейность отсутствует ($\alpha_1 = 0$), решение (21) переходит в солитон Кортевега – де Вриза

$$\eta(x,t) = a\,\text{sech}^2\left[\sqrt{\frac{\alpha a}{12\beta}}\left(x - \frac{\alpha a}{3}t\right)\right]. \quad (23)$$

Важно подчеркнуть, что полярность солитона определяется знаком коэффициента квадратичной нелинейности. Таким образом, если граница раздела находится ближе ко дну, чем к поверхности, то солитон представляет собой гребень. В противоположном случае солитон представляет собой впадину. В этом принципиальное отличие внутренних волн от поверхностных волн в однородной жидкости, когда солитон имеет только положительную полярность. Приведем также значения массы солитона

$$M_s = \int_{-\infty}^{+\infty} \eta\, dx = \sqrt{\frac{48\beta a}{\alpha}}\operatorname{sign}(\alpha), \quad (24)$$

и его энергии

$$E_s = \int_{-\infty}^{+\infty} \frac{\eta^2}{2} dx = \sqrt{\frac{16\beta a^3}{3\alpha}}. \quad (25)$$

Масса и энергия солитона уравнения Кортевега–де Вриза являются монотонными функциями его амплитуды.

*Солитон Гарднера.* Поскольку для волн в двухслойном потоке $\alpha_1 < 0$, то $0 < B < 1$. В результате, существует только одно семейство солитонов, полярность которых по прежнему определяется знаком коэффициента квадратичной нелинейности. Форма солитона для $\alpha = 1$ и $\alpha_1 = -1$ показана на рис. 1. Для малых амплитуд ($B \to 1$), солитон (21) трансформируется в солитон Кортевега – де Вриза (23). Возрастая, амплитуда волны приближается к максимальной величине

$$a_{cr} = \frac{\alpha}{|\alpha_1|}. \quad (26)$$

В этом пределе ($B \to 0$), ширина солитона стремится до бесконечности, и солитон становится так называемым «толстым» или «столообразным» солитоном. Приведем здесь также значения массы и энергии солитона уравнения Гарднера

$$M_s = 4\sqrt{\frac{6\beta}{|\alpha_1|}}\operatorname{arcth}\left(\sqrt{\frac{1-B}{1+B}}\right), \quad (27)$$

$$E_s = \frac{\alpha\sqrt{6\beta|\alpha_1|}}{\alpha_1^2}\left\{2\operatorname{arcth}\left(\sqrt{\frac{1-B}{1+B}}\right) - \sqrt{1-B^2}\right\}. (28)$$

Масса и энергия этого солитона возрастают с увеличением его амплитуды и в пределе толстого солитона стремятся к бесконечности пропорционально его ширине.

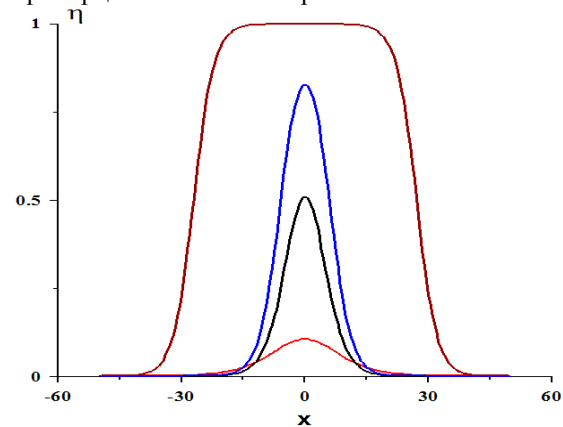

Рис.1 Солитоны уравнения Гарднера в двухслойном потоке

## АДИАБАТИЧЕСКАЯ ТРАНСФОРМАЦИЯ СОЛИТОНА ВНУТРЕННИХ ВОЛН В ПОТОКЕ ПЕРЕМЕННОЙ ГЛУБИНЫ

Если глубина потока меняется очень медленно, то волна локально является стационарной и описывается решениями, приведенными в выше. Изменения параметров солитона можно найти, используя асимптотическое решение уравнения Гарднера с переменными коэффициентами (9). Легко показать, что амплитуда солитона находится из сохранения второго (энергетического) интеграла (14), означающего сохранение потока энергии. В виду ограниченности солитона в пространстве и отсутствия отражения волны от откоса, очевидно, что второй интеграл сводится к сохранению энергии солитона в пространстве. Поэтому мы можем работать сразу с энергетическими характеристиками солитона в соответствие с формулами (25) и (28). Рассмотрим здесь трансформацию солитона малой амплитуды в рамках уравнения Кортевега – де Вриза. Учитывая выражения для коэффициентов уравнения Кортевега – де Вриза (16) и (18)

из интеграла (25) находим искомую формулу для амплитуды солитона

$$\frac{a(x)}{a_0} = \left[\frac{h_{20}}{h_2(x)}\right]^{2/3}\left[\frac{h_1 - h_2(x)}{h_1 - h_{20}}\right]^{1/3}. (29)$$

где $a_0$ – амплитуда солитона в точке с глубиной $h_{20}$. Пусть, например, солитон движется в жидкости с монотонно убывающей глубиной, при этом начальная толщина нижнего слоя превышает толщину верхнего слоя. В этом случае солитон имеет отрицательную полярность (впадина). Он существует до тех пор, пока толщины слоев не сравниваются, и в критической точке обращается в нуль, как и коэффициент квадратичной нелинейности. Характер изменения амплитуды солитона зависит от величины параметра $q = h_1/h_{20} < 1$. Исследуя на экстремум функцию (29), легко найти глубину, когда амплитуда солитона достигает максимума

$$h_2^* = 2h_1. \qquad (30)$$

Величина максимального усиления солитона есть

$$\frac{a_*}{a_0} = \left(\frac{h_{20}}{2h_1}\right)^{2/3} \left(\frac{h_1}{h_{20}-h_1}\right)^{1/3}. \qquad (31)$$

Итак, если $h_{20} < 2 \cdot h_1$ ($q > 0.5$), то амплитуда солитона только затухает при приближении к критической точке. Если же начальная толщина нижнего слоя достаточно большая, то амплитуда солитона сначала возрастает, достигая экстремального значения (31), а затем убывает до нуля. На рис. 2 приведена зависимость максимального коэффициента усиления от параметра $q$. Усиление возрастает, если глубина нижнего слоя существенно превышает глубины верхнего слоя.

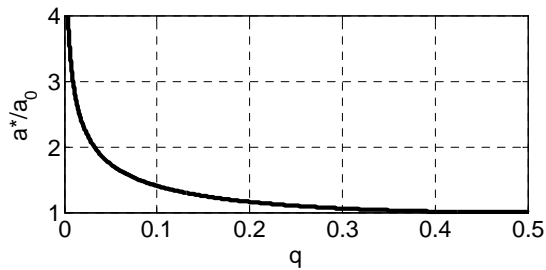

Рис.2 Зависимость максимального коэффициента усиления от параметра $q$

На рис. 3 приведена зависимость амплитуды внутренней волны в двухслойном потоке от глубины нижнего слоя, построенная по формуле (29).

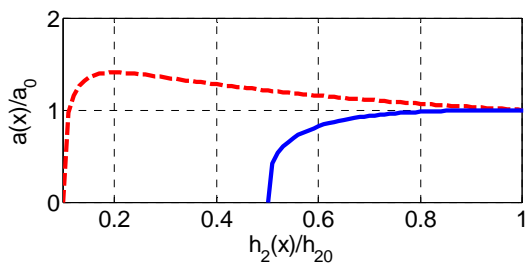

Рис.3 Изменение амплитуды внутренней волны в двухслойном потоке переменной глубины с узким верхним слоем (штриховой линией отмечен случай $h_1/h_{20} = 0.1$, а сплошной $h_1/h_{20} = 0.5$)

В критической точке в рамках асимптотического подхода амплитуда солитона обращается в нуль. На самом деле, при этом длина солитона резко возрастает, и нарушается условие медленности изменения параметров солитона. Поведение солитона в критической точке (без относительно к данной задаче) было исследовано в работах (Knickerbocker and Newell, 1980; Талипова и др., 1997). Солитон не разрушается полностью, и из него в дальнейшем рождается солитон противоположной полярности (гребень).

В случае же, если начальная толщина нижнего слоя меньше толщины верхнего слоя, то солитон изначально имеет положительную полярность (гребень) и по мере уменьшения глубины амплитуда солитона только возрастает (рис. 4). На последней стадии, когда $h_2 \to 0$, амплитуда солитона растет как $h_2^{-2/3}$

$$\frac{a(x)}{a_0} \approx \left[\frac{h_{20}}{h_2(x)}\right]^{2/3} \left[\frac{h_1}{h_1-h_{20}}\right]^{1/3}. \qquad (32)$$

Уже из этой формулы видна слабая зависимость амплитуды солитона от отношения $h_{20}/h_1$, которая и заметна на рис. 4. Отметим также отличие в законах изменения амплитуд внутреннего и поверхностного солитона, для поверхностных волн $a \sim h^{-1}$ (Пелиновский, 1996).

В точке, когда $h_2 = 0$, двухслойный поток трансформируется в однослойный, и внутренние волны при этом невозможны. Физически это связано с разрушением солитона в переходной зоне (по аналогии с поверхностными волнами такую зону можно назвать зоной наката внутренних волн), где его амплитуда резко возрастает. Этот процесс необходимо исследовать в рамках более полных по нелинейности уравнений. Если амплитуда волны достаточно мала, то обрушение в зоне наката может и не происходить, и внутренняя волна отражается от откоса (Талипова и Пелиновский, 2011).

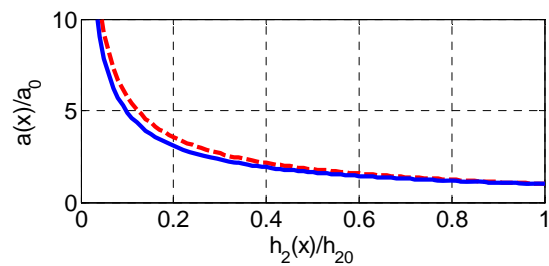

Рис.4 Изменение амплитуды внутренней волны в двухслойном потоке переменной глубины с малой толщиной нижнего слоя (штриховой линией отмечен случай $h_1/h_{20} = 2$, а сплошной $h_1/h_{20} = 6$)

**ЗАКЛЮЧЕНИЕ**

Разработанная ранее физико-математическая модель трансформации внутренних волн конечной амплитуды в стратифицированной жидкости применена для анализа уединенной волны малой амплитуды (солитона) в двухслойном потоке с уменьшающейся глубиной. Показано существование двух критических точек, в которых солитон разрушается. Первая из них возникает, когда толщины слоев сравниваются между собой, и коэффициент квадратичной нелинейности обращается в нуль. В этой точке солитон отрицательной полярности исчезает, и в дальнейшем возникает солитон положительной полярности. Другая критическая точка соответствует превращению двухслойного потока в однослойный. В таком потоке внутренние волны не могут существовать, и волны разрушаются при подходе к этой точке.